\documentclass{article}
\usepackage{waspaa23,amsmath,graphicx,url,times}
\usepackage{color, hyperref}
\usepackage{dingbat}

\title{Low bit rate binaural link for improved ultra low-latency low-complexity multichannel speech enhancement in Hearing Aids}

\name{Nils L. Westhausen and Bernd T. Meyer \thanks{This work was funded by the Deutsche Forschungsgemeinschaft (DFG, German Research Foundation) under Germany's Excellence Strategy – EXC 2177/1 - Project ID 390895286 and by - Project-ID 352015383 - SFB 1330.}}
\address{Communication Acoustics \& Cluster of Excellence Hearing4all\\
  Carl von Ossietzky University, Oldenburg, Germany
}

\begin{document}

\ninept
\maketitle

\begin{sloppy}

\begin{abstract}
\ninept
Speech enhancement in hearing aids is a challenging task since the hardware limits the number of possible operations and the latency needs to be in the range of only a few milliseconds. 
We propose a deep-learning model compatible with these limitations, which we refer to as Group-Communication Filter-and-Sum Network (GCFSnet). GCFSnet is a causal multiple-input single output enhancement model using filter-and-sum processing in the time-frequency domain and a multi-frame deep post filter. All filters are complex-valued and are estimated by a deep-learning model using weight-sharing through Group Communication and quantization-aware training for reducing model size and computational footprint. For a further increase in performance, a low bit rate binaural link for delayed binaural features is proposed to use binaural information while retaining a latency of 2~ms. The performance of an oracle binaural LCMV beamformer in non-low-latency configuration can be matched even by a unilateral configuration of the GCFSnet in terms of objective metrics.      

\end{abstract}


\ninept
\section{Introduction}
\ninept
A standard hearing-aid configuration for a person with a bilateral hearing loss normally consists of two devices for the left and right ear, respectively.
For speech enhancement in this setting, it is desirable to exploit binaural information, taking into account signals from the bilateral devices with a relatively large spacing (approximately 18~cm). 
A traditional approach to use this information is binaural beamforming such as the binaural minimum variance distortionless response (BMVDR) \cite{BMVDR} or the binaural linearly constraint minimum variance (BLCMV) \cite{LCMV} beamformer, both of which are able to preserve spatial cues.
Recently, deep-learning based approaches for multichannel speech enhancement were proposed \cite{low_latency, waspa_09_subband, fasnet}. 
Only few investigate causal low-latency processing  \cite{marvin_2022, convtasnet_bin, low_latency} as required for hearing aids and are implemented as low-complexity algorithms \cite{kovalyov2023dfsnet, lowCompLowLatecy}, or take into account binaural signals \cite{marvin_2022, convtasnet_bin}, which is beneficial for spatial cue preservation. 

For interlinked bilateral hearing aids, a transmission of microphone signals between devices is required.
The binaural algorithms mentioned earlier \cite{BMVDR, LCMV, marvin_2022, convtasnet_bin} assume synchronized microphone signals.
However, inter-device transmission introduces a delay since the microphone signals of the ipsilateral ear must be properly delayed by the transmission delay of the microphone signals of the contralateral ear.
This strategy increases the overall latency, which results in a degradation of the perceived sound quality \cite{Stone2008}.
Further, the bandwidth of transmission (which should work wirelessly since a wired connection between hearing aids is not desirable) can be limited \cite{Derleth2021}. 
This potentially reduces the quality of the transmitted signals which impacts the audio quality of algorithms using all microphone signals.

In this paper, we propose a system that considers both challenges, latency and limited bandwidth, of the binaural transmission.  
Our investigation is based on two low-latency multiple input single output (MISO) deep-speech enhancement models, one running on each side on the behind-the-ear devices. The approach is based on deep filter-and-sum beamforming in the spectral domain with a multi-frame post filter. 
The model is using weight sharing and group-communication (GC) \cite{groupcomm} for a reduced size and computational footprint. Additionally, quantization-aware training is applied \cite{quant_aware} for potentially running the model on embedded devices. The size and complexity of the model was chosen to be compatible with next-gen hearing aids SOCs \cite{smartHeap}.
For connecting the left and right device, we are using binaural features delayed by the transmission delay, while spatial filtering is only performed unilaterally so it is not influenced by the transmission delay and degradation due to limited bandwidth. During training, the latency and bit rate of the transmitted microphone signals is varied to simulate a transmission such that the model is robust against these factors.
The approach is evaluated with the Scale Invariant Signal-to-Distortion Ratio (SI-SDR), the Perceptual Evaluation of Speech Quality (PESQ), the better-ear Hearing Aid Speech Perception Index (HASPI) and the Modified Binaural Short-Time Objective Intelligibility (MBSTOI).
\section{Methods}
\vspace{-1mm}
\subsection{Deep spatial and post filtering}
\vspace{-1mm}
The proposed approach is designed for enhancement in acoustic scenes with one target speaker, interfering speakers and a diffuse noise source.
It is based on the following observations: 
The multichannel STFT representation of the mixture $y$ can be written as
\begin{equation}
    Y(m, t, f) = X_{S}(m,t,f) + \sum_{i=1}^{I} X_{V_{i}}(m,t,f) + X_{D}(m,t,f) 
\end{equation}
$Y$ is the mixture while $m$, $t$, $f$ are the microphone, frame and frequency index, respectively. $X_{S}$ corresponds to the reverberant complex time frequency (TF) representation of the anechoic target speech signal $s(n)$. $X_{V_{i}}$ is the reverberant TF-representation of the interfering speaker $v_i(n)$ with index $i$. The reverberant TF-representation of the noise $d(n)$ is represented by $X_{D}$

The aim is to extract the direct speech signal $s(n)$ for the left and the right side separately.
The left $l$ and right $r$ channels are each estimated with complex filter-and-sum beamforming using only microphone signals from the corresponding side
\begin{equation}
\label{eqn:filter_channel}
\Tilde{S}(t, f) = \sum_{m=1}^{M} Y(m, t, f) \cdot W(m, t, f)
\end{equation}
where $\Tilde{S}$ denotes an estimated intermediate TF-representation of the target speaker and $W$ denotes the complex filter weights estimated by the model. $M$ represents the number of microphones on each side.
Subsequently, a causal single-channel multi-frame post-filter of length $K$ is applied to the left and the right channel individually: 
\begin{equation}
\label{eqn:filter_post}
\hat{S}(t, f) = \sum_{k=0}^{K} \Tilde{S}(t - k, f) \cdot C(t,k,f).
\end{equation}
$C$ denotes the complex multi-frame filter. The multi-frame postfilter is inspired by \cite{deepfilternet2}.
$\hat{S}(t, f)$ is transformed back to the time-domain by an iSTFT. The general algorithm structure is illustrated in Fig.\ref{fig:filter-structure}.

\begin{figure}[t]
  \centering
  \includegraphics[width=0.85\linewidth]{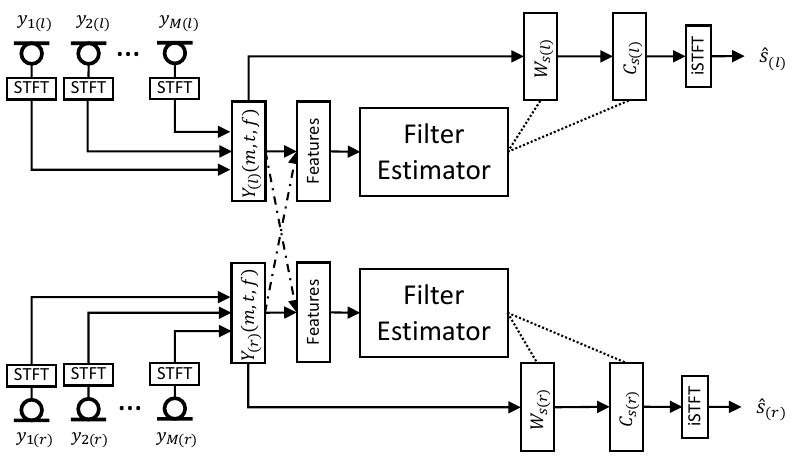}
  \caption{Illustration of the proposed approach for spatial filtering with post filtering. The filters for the left and right side are estimated by separate models. The dashed-dotted line symbolizes
  an exchange of the microphone signals with transmission delay preceding feature calculation. There are $M_l$ microphones on the left and $M_r$ microphones on the right available.}
\label{fig:filter-structure}
\vspace{-4mm}
\end{figure}
\vspace{-2mm}
\subsection{Architecture}
\vspace{-1mm}
The complex filters for the left and right side are estimated by two models with an identical architecture.
The architecture is based on grouping and weight-sharing across groups. 
The model is trained with quantization-aware training \cite{quant_aware} where, when not noted otherwise, the weights of the layers are quantized with 8~bits and the bias is quantized with 16~bits. The inputs and outputs to the layers are quantized with 16~bits so it simulates fixed-point calculations. A linear fixed-range quantization with range [-1,1] is used with a quantization equalizer (quantEQ) as proposed in \cite{pruning} to reduce the risk of information loss from overflow implemented by a Depthwise-Conv (D-Conv) with kernel size 1. The weights of this D-Conv are not quantized.
All layers are using a $tanh$ activation function for bound latent representations.

After the scaling with the quantEQ, the input to the model of shape $(t,B)$ is projected by a fully connected (FC) layer to an intermediate representation of size $P$. The intermediate projection is split into $G$ parallel groups. The grouping mechanism is inspired by \cite{groupcomm}.

Next the groups are mapped to the hidden size $U$ by a shared FC layer. Subsequently, the groups are processed by a shared conv-module containing causal depthwise-separable convolutions (DS-CONV) with kernel size 5 and 3 and an additive skip connection around the DS-Conv layers implemented by an D-Conv layer. Adding a D-Conv to the skip connection is inspired by \cite{deepfilternet2}. 

The conv module is followed by Transform Average Concatenate (TAC) \cite{fasnet_tac} for GC \cite{groupcomm} with a hidden dimension of $2U$. The PReLU activation in the original TAC is changed to $tanh$ for bound representations. 
The next shared module contains 2 stacked GRU layers with $U$ units and as well an additive skip connection implemented by a D-Conv. In the following the groups are again processed by TAC.
The conv module is chosen to account better for short-term dependencies while the module containing the GRUs focuses more on long-term dependencies. The weights of the conv and GRU module are shared between groups as well as between models.
For ungrouping, the hidden representations of the groups are mapped to size $P/G$ by a shared FC layer. The hidden representation are combined to form the final representation of size $P$.
The real and complex part of the spatial and post filters are estimated by FC layers
from the final representation.
The architecture is visualized in Fig.~\ref{fig:filter-network}.
\begin{figure}[t]
  \centering
  \includegraphics[width=1.0\linewidth]{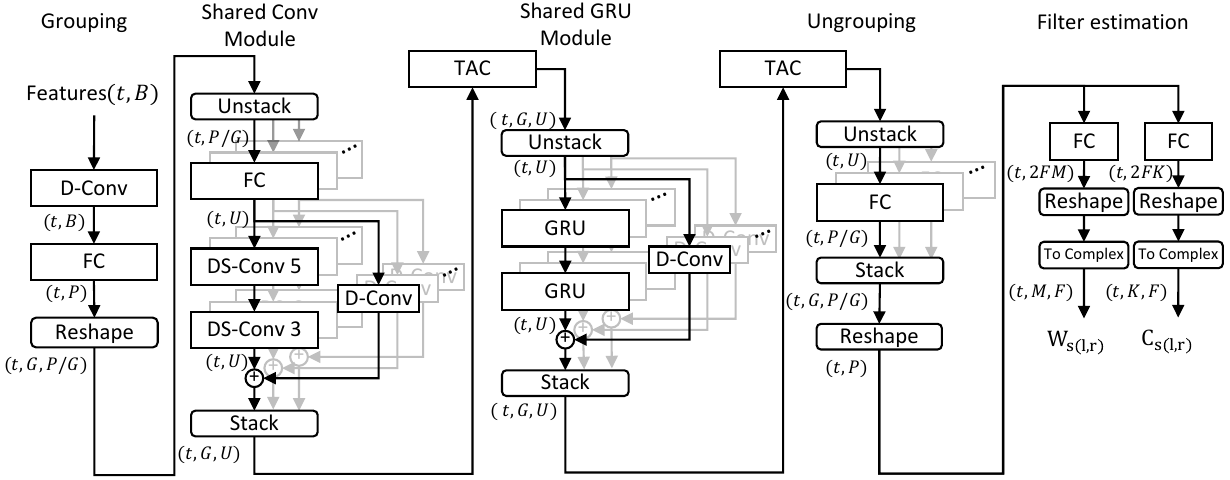}
  \caption{Illustration of the proposed filter estimation model. }
\label{fig:filter-network}
\vspace{-4mm}
\end{figure}
\vspace{-2mm}
\subsection{Features and simulation of low bit rate binaural link}
\vspace{-1mm}
In total three different configurations for the features are considered depending on the availability of microphone signals. The base configuration is the unilateral scenario ($\mathrm{GCFSnet_{uni}}$) without any communication between the left and right side.
In this case, each side has two microphone signals available.
The log magnitude of the STFT of each channel and the sin and cos phase difference (IPD) between both microphones serve as features, which are concatenated to form the feature vector of size B. The IPD features are inspired by \cite{convtasnet_bin}. 
The log compression and the bound range of sin and cos are advantageous when targeting quantized models with a bound dynamic range compared for instance to the raw real and imaginary part of the STFT.

In the second case, a transmission of the microphone signals between the the left and right side is available ($\mathrm{GCFSnet_{lowB}}$). It is assumed that the transmission introduces a known delay and has a possibly variable or lower bit rate.
Initially, features are calculated as in the unilateral scenario. The first \emph{binaural} feature added is the log magnitude of the STFT of the transmitted signals which is concatenated to the feature vector. The second part of the binaural features are binaural sin(IPD) and cos(IPD).
To calculate binaural IPDs, the ipsilateral signals are delayed by the known transmission time. Subsequently, the IPDs between the reference microphone channel and all other microphones are calculated. In case of four microphones, we obtain three IPD vectors. The sin(IPD) and cos(IPD) are appended to the feature vector. 
The advantage of this delayed feature scheme is that it is still possible to use the increased localisation performance of the binaural signals to certain degree while not impacting the processing delay. 
 
For the third case, a binaural transmission with no latency and fixed bit rate is assumed ($\mathrm{GCFSnet_{bin}}$). The log magnitude of all microphones is concatenated together with the sin and cos IPDs between the reference microphone channel and all other microphone channels, as described above.

The binaural transmission during training and testing is simulated by delaying and quantizing the microphone signals of the contralateral sides. Quantization is performed as fake-quantization in the same way as for quantization-aware training with a range between -1 and 1. This procedure assumes a uncompressed audio transmission at a certain bit rate. The delay is implemented as a front zero padding where the delay is always assumed to be an integer factor of the frame-shift of the STFT.

During training, a variable random delay between 4 to 12~ms and random bit rate between 4 and 16~bits is applied. The delay range is chosen based on experience with near field magnetic induction (NMFI) for microphone signal transmission.
The testing is always performed with static delay and static bit rate for all utterances. The delay is considered as known.
\begin{figure*}[t]
  \centering
  \includegraphics[width=1.0\linewidth, trim={0 0 0 0},clip]{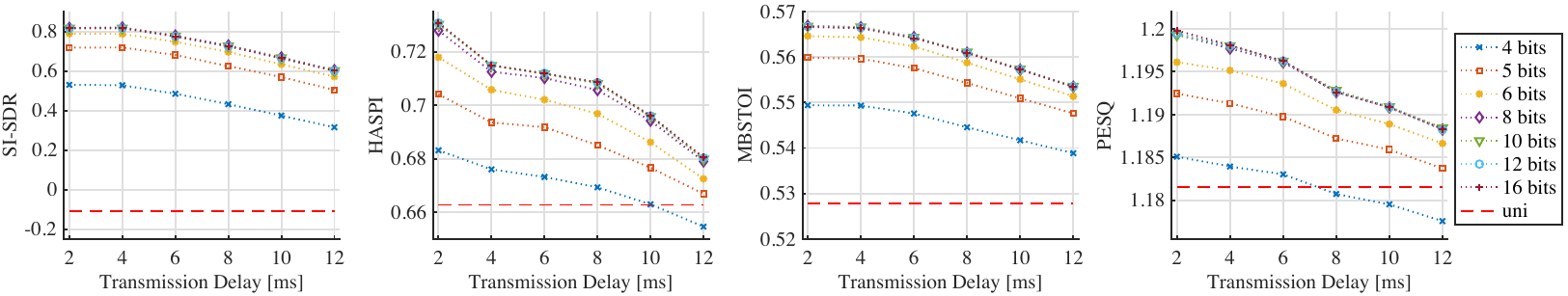}
  \caption{Results for $\mathrm{GCFSnet_{lowB}}$ tested on multiple delays and bit rates. The red dashed line shows the result for $\mathrm{GCFSnet_{uni}}$ as reference.}
\label{fig:result_plot}
\vspace{-4mm}
\end{figure*}
%
%
%
\vspace{-2mm}
\subsection{Dataset and acoustic scene generation}
\vspace{-1mm}
The basis for training, evaluation and test data are the min train-360 (146h), min dev (11h) and min test (11h) split of the Libri3Mix Corpus \cite{cosentino2020librimix} with WHAM! noise \cite{Wichern2019WHAM} at a sampling rate of 16~kHz.
This corpus was chosen since it is publicly available and contains a large variety of speakers and ambient noise.
60k binaural room impulse responses (BRIR) are simulated by RAZR \cite{wendt2014a} using the fixed high-resolution HRTF set from \cite{Thiemann2019} of the B\&K HATS with behind-the-ear hearing-aid dummies with three microphones on each side (front, mid and rear) for multichannel mapping.
RAZR is able to return BRIR $H$ as well as the parts of the impulse response were $H = H_{direct} + H_{early} + H_{late}$. $H_{direct}$ is chosen as a properly delayed training target to include dereverberation. At the same time, this signal is used as a clean target for the objective measures.
Since we explore \emph{binaural} scenarios with a target speaker from the front without \emph{collocated} interfering speakers, existing datasets (e.g., \cite{graetzer21_interspeech}) are not applicable; hence, we created a new data set from commonly available resources. 

The acoustic scene used for scenario contains a receiver, a target speaker, two interfering speakers and a diffuse noise source.
Room parameters were drawn from equal distributions ranging from 0.1 to 0.5~s for the RT60, 2.5 to 4.5~m for the room height, 3 to 10~m for the room widths and 12 to 100~$\mathrm{m}^2$ for the surface area. 
The receiver is positioned at random up to 1~m from the center of the room at a height between 0.9 to 1.8~m.
All sources are at least 1~m away from the walls. 
The target speaker is located in the front between -12° to 12° azimuth, between 0.75 to 1.5~m from the receiver and at a height between 0.9 to 1.8~m. 
The interfering speakers are placed at any angle omitting -15° and 15° azimuth with a minimum 10° angular distance from the target speaker. The distance to the receiver is between 0.75 and 2~m and the height is as well between 0.9 to 1.8~m. The diffuse noise is modeled by the output of the feedback delay network ($H_{late}$) of RAZR for a source positioned at a random position at least 1~m away from the receiver. The goal of this configuration is to learn a clear steering towards speech signals from the front, motivated by static beamformers directed to the front that are often used in current hearing aids\cite{Derleth2021}.
Our intuition for the learned static configuration is that it could be relatively robust and without depending on additional information such as direction of arrival or speaker identity. 

During training, the training samples are created online. The mixing of signals is performed as in \cite{graetzer21_interspeech}. The better-ear SNRs of the interfering speech signals relative to the target speaker are drawn from a normal distribution $\mathcal{N}(0, 4.1^2)$~dB and for the noise from the normal distribution $\mathcal{N}(6.2, 4.4^2)$~dB. These distributions are corresponding to the original distribution of LibriMix data. The level of the mixture is drawn from $\mathcal{N}(-26, 5^2)$~dB~FS. The test set is created offline and uses the original SNRs and the original input scales of Libri3Mix.
\vspace{-2mm}
\subsection{Training configuration}
\vspace{-1mm}
The model is trained for 100 epochs with a batchsize of 16. ADAM is used as optimizer with an initial learning rate of 1e-3. The learning rate is multiplied by 0.98 every two epochs. If the loss on the validation set does not decrease for 5 consecutive epochs, the learning rate is multiplied by 0.8. The weights of the model are saved when the loss on the validation set decreases.
For gradient clipping, we use AutoClip \cite{autoclip} with $p=10$ for smoother training and better generalization.
The training setup uses TensorFlow 2.11.
The frame length of the filtering framework is 2~ms with a shift of 1~ms with FFT-length of 64 using equal front and back zero padding.
A $\sqrt{Hann}$ window is applied for calculating STFT and iSTFT.
$P$ is set to 128, $U$ to 32 and $G$ to 8 for all configurations. $K$ is set to 5 to capture some inter-frame correlations. This results in 164k parameters and approximately 0.36 GMAC/S for a single model on one side. The parameters were chosen so that the theoretical size and complexity would fit the capacity of \cite{smartHeap}.
The front and the rear microphone from each side are used as input for all models, with the signals of the front microphone being used as reference. The left and the right model are always trained at the same time.

The training uses an equally-weighted combination of the compressed spectral MSE (cMSE) \cite{braun_loss} and the phase-constrained magnitude loss (PCM) \cite{pca_loss} as loss function. We noticed a slight improvement using both loss functions compared to using either one of them alone. We assume that the slightly different properties of the loss functions are complementary. The loss functions are calculated after signal reconstruction with a window length of 20~ms a shift of 10~ms and an FFT-length of 320.

We use an oracle BLCMV beamfomer as baseline with a block length of 64~ms and a block shift of 32~ms. The beamformer has access to all reverberant components of the mixture for estimating the relative transfer functions of the speech sources. The interfering speakers are applied with a gain of -40 dB. We chose this baseline because of its optimal binaural processing and high signal quality.
We did not use any neural baseline since the goal of this paper is to study the effect of the binaural link with the GCFSnet. 
\vspace{-1mm}
\subsection{Evaluation metrics}
\vspace{-2mm}
For evaluation, SI-SDR \cite{SI-SDR}, PESQ \cite{pesq} the better-ear HASPI \cite{HASPI} (Range: 0 to 1) and MB-STOI \cite{MBSTOI} (Range: 0 to 1) are considered. For SI-SDR and PESQ, the mean over the the binaural signal is calculated.
HASPI is used in normal hearing configuration and the mixture is scaled to represent a playback level of 65~dB SPL.
The reference for all metrics is the target speech signal convolved with the direct part of the BRIR.
\section{Experiments and results}
The results in terms of objective metrics are shown in Tab.~\ref{tab:results}, while results for different bit rates and transmission delays are shown in Fig.~\ref{fig:result_plot}. 

\textbf{General performance}:
All GCFSnet configurations show a clear improvement in terms of objective measures compared to the unprocessed condition. The $\mathrm{GCFSnet_{bin}}$ performs better than $\mathrm{GCFSnet_{uni}}$, which is expected since $\mathrm{GCFSnet_{bin}}$ uses fully synchronized microphone signals for the left and right devices. $\mathrm{GCFSnet_{lowB}}$ can increase the metrics compared to $\mathrm{GCFSnet_{uni}}$. This effect is most prominent for the SI-SDR. For the other metrics, a slight improvement can be observed. 
When comparing the GCFSnet configurations with the oracle BLCMV beamformer $\mathrm{GCFSnet_{uni}}$ already matches the performance for HASPI and outperforms the BLCMV for SI-SDR and PESQ. The low scores of the BLCMV for SI-SDR can be explained by the lack of explicit handling for dereverberation of the BLCMV used here. This results in more preserved reverberation which influences the SI-SDR score. 
For MBSTOI, the BLCMV shows the highest scores together with $\mathrm{GCFSnet_{bin}}$ at 8 and 16~bits. This indicates that $\mathrm{GCFSnet_{bin}}$ and the BLCMV are better in preserving binaural information important for MBSTOI. However, these improvements increase latency, since they require synchronized microphone signals.      
\begin{table} [th]
\vspace{-5mm}
  \caption{Results in terms of objective metrics. (Delay in ms)}
  \label{tab:results}
  \centering
  \footnotesize
  \begin{tabular}{l@{\hspace{0.2cm}} c@{\hspace{0.2cm}} c@{\hspace{0.2cm}} c@{\hspace{0.2cm}} c@{\hspace{0.2cm}} c@{\hspace{0.2cm}} c@{}}
    \hline
   \textbf{System}  & \textbf{Delay} &  \textbf{Bits} & \textbf{SI-SDR} & \textbf{HASPI} & \textbf{MBSTOI} & \textbf{PESQ}\\
    \hline
    Unprocessed   & & &  -9.20   &   0.17    &  0.38  & 1.06\\
    \hline
    BLCMV & & & -1.52   &    0.67   &  0.60  & 1.11 \\
    \hline
    $\mathrm{GCFSnet_{uni}}$ &  & & -0.11    &   0.66   & 0.53 & 1.18\\
    \hline
    $\mathrm{GCFSnet_{lowB}}$ & 12 & 6 & 0.57    &    0.67  &  0.55 & 1.19   \\
    $\mathrm{GCFSnet_{lowB}}$ & 12 & 8 & 0.60    &    0.68  &  0.55 & 1.19   \\
    $\mathrm{GCFSnet_{lowB}}$ & 12 & 16 & 0.60    &    0.68  &  0.55 & 1.19   \\
    \hline
    $\mathrm{GCFSnet_{lowB}}$ & 6 & 6 & 0.75    &    0.70  &  0.56 & 1.19   \\
    $\mathrm{GCFSnet_{lowB}}$ & 6 & 8 & 0.79    &    0.71  &  0.56 & 1.20   \\
    $\mathrm{GCFSnet_{lowB}}$ & 6 & 16 & 0.77    &    0.71  &  0.56 & 1.20   \\
    \hline
    $\mathrm{GCFSnet_{bin}}$ & 0 & 6 &   1.44  &  0.73   & 0.59  &  1.22  \\
    $\mathrm{GCFSnet_{bin}}$ & 0 & 8 &   1.51  &   0.75  &  0.59 &  1.23  \\
    $\mathrm{GCFSnet_{bin}}$ & 0 & 16 & 1.45    &   0.75  &  0.59 & 1.23   \\
    \hline
  \end{tabular}
  \vspace{-1mm}
\end{table}
\ninept

\textbf{Effect of delay and bit rate}: Fig.~\ref{fig:result_plot} shows the results for an evaluation of $\mathrm{GCFSnet_{lowB}}$ with different delays and bit rates. In case of SI-SDR, all evaluated delays and bit rates are performing better than the $\mathrm{GCFSnet_{uni}}$. This is interesting, since it implies that even for 4~bits where only 16 quantization steps are available, the model can use the binaural information to some degree to increase the SI-SDR. A similar result is observed for MB-STOI, while for HASPI and PESQ at least 5~bits are required for a more constant improvement over $\mathrm{GCFSnet_{uni}}$. For a bit rate over 6~bits, performance saturates for all metrics. We conclude that transmissions at a relative low bit rate can improve noise suppression performance. Another clear trend for all metrics is the reduction in improvement with increasing delay, which is intuitive since the delayed binaural information gets less relevant for the current frame.
At 4~bits the HASPI score for 12~ms delay and PESQ scores above 6~ms delay are even lower than for $\mathrm{GCFSnet_{uni}}$. This means at lower bit rates the link can introduce distortions impacting HASPI and PESQ.
In general the magnitude of the improvements over $\mathrm{GCFSnet_{uni}}$ are not very large (except for Si-SDR), but on the other hand any improvement in speech enhancement performance can be beneficial for intelligibility.  

\textbf{Effect of binaural features}: Another interesting aspect is the importance of the two parts of the binaural features, the log-magnitude STFT of the transmitted microphone signals and the sin(IPD) and cos(IPD). For this reason we compare a model only trained with additional binaural log-magnitude and one restricted to additional binaural sin(IPD) and cos(IPD). The results are shown in Tab.~\ref{tab:results_2}. The model limited to binaural IPD features is slightly better than the model only using binaural log-magnitude.
This effect is most prominent for SI-SDR, indicating that the direction of arrival information embedded in the IPD features is more relevant than the spectral features of the log-magnitude.
The model using both features still performs better for SI-SDR, HASPI and PESQ. 
\begin{table}[th]
\vspace{-4mm}
  \caption{Evaluation of the influence of binaural features. Models are tested with a delay of 6~ms and a bandwidth of 8~bits.}
  \label{tab:results_2}
  \centering
  \footnotesize
  \begin{tabular}{l@{\hspace{0.2cm}} c@{\hspace{0.2cm}} c@{\hspace{0.2cm}} c@{\hspace{0.2cm}} c@{\hspace{0.2cm}} c@{\hspace{0.2cm}} c@{}}
    \hline
   \textbf{System} & \textbf{LogMag} & \textbf{IPD} & \textbf{SI-SDR} & \textbf{HASPI} & \textbf{MBSTOI} & \textbf{PESQ}\\
    \hline
    $\mathrm{GCFSnet_{lowB}}$ & \checkmark & \checkmark & 0.78    &    0.71  &  0.56 & 1.20   \\
    $\mathrm{GCFSnet_{lowB}}$ &  & \checkmark & 0.66    &    0.69  &  0.56 & 1.19   \\
    $\mathrm{GCFSnet_{lowB}}$ & \checkmark &   & 0.06    &    0.67  &  0.53 & 1.18   \\
    \hline
  \end{tabular}
  \vspace{-2.5mm}
\end{table}
\ninept

\textbf{Future work}: In future work, more dynamic scenarios should be considered for evaluation and/or for training. Also a steering similar to \cite{tesch2022spatially} would be desirable. Another interesting aspect would be the comparison of different spatial and binaural features for delayed binaural features such as the normalized cross correlation features suggested in \cite{fasnet_tac}. 
Another important step is the evaluation of this algorithm with hearing-impaired subjects, BTE devices and real-time audio processing. For this task, the algorithm could be integrated as a plugin into open-source research platforms such as the open master hearing aid \cite{openMHA}.  
Finally, the proposed system could be ported to a fully fixed-point C implementation to enable testing on a platform such as \cite{smartHeap}. 
This platform also includes near-field magnetic induction for the transmission of audio signals.  

\section{Conclusion}
This study introduced a low-latency and low-complexity deep-speech enhancement model for hearing aids performing filter-and-sum beamforming and postfiltering in the spectral domain.
The results show that a limited low-bitrate binaural link can improve speech enhancement performance while not increasing the system latency by the transmission delay between devices. The approach is able even to match the performance of an oracle BLCMV beamformer with a high-latency configuration.

\bibliographystyle{IEEEtran}
\bibliography{2023_WASPAA_PaperTemplate_Latex}

\end{sloppy}
\end{document}